\documentstyle[aps,prb,tabularx,graphicx,twocolumn]{revtex}

\def\eqref#1{(\ref{#1})}

\begin{document}

\wideabs{

\title { A Modified Random Phase
Approximation of Polyelectrolyte Solutions }

\author{A. V. Ermoshkin, M. Olvera de la Cruz}

\address{Department of Materials Science and Engineering,\\
Northwestern University, Evanston, IL, USA 60208-3108}

\date{\today}

\maketitle

\begin{abstract}

We compute the phase diagram of salt-free polyelectrolyte solutions using a 
modified Debye-Huckel Approach. 
We introduce the chain connectivity via the Random Phase 
Approximation with two important modifications. We modify the electrostatic 
potential at short distances to include a bound on the electrostatic 
attractions at the distance of closest approach between charges. This 
modification is shown to act as a hard core in the phase diagram of electrolyte 
solutions. We also introduce a cut-off on the integration of the modes of wave 
length smaller than the size over which the chains are strongly perturbed by 
the electrostatic interactions.  This cut-off is shown to be essential to 
predict physical phase diagram in long chain solutions. 
\end{abstract}

}

\section{Introduction}
Debye-H$\rm\ddot u$ckel (DH) theory derived nearly 80 years ago
successfully explains many thermodynamic properties of dilute
symmetric electrolyte solutions.\cite{book} Recent computer
simulations of hard-sphere ionic fluids\cite{Panagio02}
show that the DH approach for finite size ion solutions provides a
rather accurate estimate of the critical temperature, although it
underestimates the critical density. A conceptually simple
modification of the DH approach, based on accounting for the
interactions of ionic pairs introduced by Bjerrum\cite{Bjerrum}
with residual ionic fluid, however, gives better estimates for the
critical parameters.\cite{Fisher93}

Though a large scientific effort has been made towards
understanding the thermodynamics of simple electrolyte solutions,
flexible charged chains solutions are not understood. The Random
Phase Approximation (RPA), a DH approach which includes the
structure function of the chains, is widely used to describe
polyelectrolytes solutions.\cite{Borue88,Joanny90} RPA, however,
provides unphysical coexisting curves in salt free polyelectrolyte
solutions.\cite{Mahdi00} In particular, RPA predicts that the
critical density decreases with increasing the chains degrees of
polymerization $N$, and it goes to zero as $N$ goes to infinite.
Moreover, the critical temperature increases rapidly as $N$
increases suggesting that strongly charged chains are not water
soluble in salt-free solutions. Recent computer simulations of
charged chains reveal a critical density rather insensitive to the
value of $N$, and a critical temperature that does increases as
$N$ increases but not as rapidly as predicted by the RPA
approach.\cite{Kumar02} A one component plasma approach to
polyelectrolytes solutions gives both a critical density and a
critical temperature insensitive to the the degree of
polymerization $N$.\cite{Muthu02} The one component plasma
approach ignores the contribution from the fluctuations of the
monomers. That is, the one component plasma assumes that the
chains provide a constant charge density support over which the
counterions fluctuate. It is difficult to demonstrate and
understand why and when it is possible to neglect the chain
density fluctuations. Moreover, the value of the critical
temperature is strongly underestimated in the one component plasma
approach to polyelectrolytes salt-free solutions.

Here we describe a modified polyelectrolyte RPA model that
predicts realistic values of the critical temperature, and a
critical density rather insensitive to $N$ in agreement with the
simulations. The RPA approach described here includes two
modifications. The first modification is related to the cut-off in
RPA. In salt free polyelectrolyte solution the electrostatic
contribution to the free energy of charged Gaussian chains does
not reduces to the DH limiting law.\cite{Borue88} Instead, a more
strong electrostatic contribution results in salt-free solutions
that generates instabilities in the free energy as the monomer
concentration goes to zero at room temperatures in strongly
charged chains.\cite{Mahdi00} Scaling
arguments\cite{deGennes79,Schiessel,Dobrynin99} and computer
simulations,\cite{Stevens95,Kremer} however, show that charged
chains are stretched in monovalent ionic solutions on length
scales smaller and of the order of the inverse screening length.
Therefore, it is not possible to include the monomer density
fluctuations within RPA, which assumes that the electrostatic
interactions do not perturbed the chain statistics (a linear
response theory). Here we introduce a concentration dependent cut
off proportional to the inversive screening length in the RPA
electrostatic contribution to include only the contribution from
the unperturbed Gaussian chains. We also discuss other
possibilities for the cut-off that yield similar modifications in
the phase diagram. The second and less important modification
involves including the contribution from the hard core of the ions
and the monomers. In simple electrolyte solutions the hard core of
the ions is included in the electrostatic contribution to the free
energy to avoid spurious results as the concentration of ions
increases. In this paper we modify the interaction potential and
show that our modified potential describes the polyelectrolyte
solution at the same level of approximation as the modified DH for
hard-sphere ionic fluids.

In Section II we describe the modified RPA model for salt-free
flexible strongly charged chains. In Section III we discuss the
electrostatic interaction potential used in the model. In Section
IV we state the problem of RPA describing Gaussian chains and the
required modifications. In Section V we outline and discuss the
results, and give some conclusions.

\section{Model and Free Energy}
Let us consider a mixture of negatively charged polyelectrolyte
chains and a residual ionic fluid. We assume that every chain
consists of $N$ monomers each carrying a negative charged $-q$ and
we denote the concentration of monomers as  $\rho_-$. Every ion in
the system carries the charge $+q$ and their total concentration
is $\rho_+$. Due to electroneutrality the following condition
holds
\begin{equation}
\rho_-=\rho_+
\label{electronutrality}
\end{equation}
We write free energy of the system in the form
\begin{equation}
F=F_{\rm ref}+F_{\rm el}
\label{F}
\end{equation}
Here $F_{\rm ref}$ is the free energy of the system without any electrostatic interactions
and in the limit of good solvent it could be written as
\begin{equation}
\frac{F_{\rm ref}}{TV}=\rho_+\ln\rho_++\frac{\rho_-}{N}\ln\rho_-+
\frac{1}{a^3}(1-\rho^*)\ln(1-\rho^*)
\label{Fref}
\end{equation}
where $T$ and $V$ is the temperature (in units of Boltzmann
constant $k_B$) and the volume of the system respectively. We also
assumed that monomers of the chains and dissociated ions have the
same sizes $a$, so we can define the reduced density $\rho^*$ as
follows
\begin{equation}
\rho^*=a^3(\rho_-+\rho_+)
\label{rho_star}
\end{equation}
The second term on the right hand side of Eq. \eqref{F} is the electrostatic contribution to the
free energy. Using RPA approach we write it in the form\cite{Borue88}
\begin{eqnarray}
\frac{F_{\rm el}}{TV}=\frac{1}{4\pi^2}\int\limits_0^{\cal K}
\left[\ln\left(1+(\rho_++\rho_-g(k))\frac{U(k)}{T}\right)-\right. \nonumber \\
\left.(\rho_++\rho_-)\frac{U(k)}{T}\right]k^2dk
\label{Fel}
\end{eqnarray}
where the structure function of the chain $g(k)$ has the following definition
\begin{equation}
g(k)=\frac{1}{N}\sum\limits_{i,j}\left<e^{-i{\bf k}({\bf r}_i-{\bf r}_j)}\right>
\label{gkdef}
\end{equation}
Indices $i$ and $j$ in Eq. \eqref{gkdef} run over all monomers of
the chain and the average $<...>$ is taken  over all possible
chain confirmations. $U(k)$ in Eq. \eqref{Fel} is the Fourier
transform of the interaction potential between charges and the
parameter $\cal K$ is used to cut the integration over the large
values of $k$. The purpose of the present work is to discuss the
RPA approximation, and determine appropriate forms of $U(k)$ and
$\cal K$.

\section{Interactions between Charges}
For dilute solutions of electrolytes the potential of interaction between charges can be taken in
the form
\begin{equation}
U(r)=\frac{q^2}{r\varepsilon}
\label{Ur}
\end{equation}
where $r$ is the distance between charges and $\varepsilon$ is the
dielectric constant of the media. With this potential Eq.
\eqref{Fel} calculated for a simple case of $N=1$ and $\cal
K=\infty$ reduces to $-\kappa^3/12\pi$. That is, the electrostatic
contribution reduces to the limiting case ($\kappa a\ll1$) of the
DH expression for the free energy of hard-sphere ionic fluid,
\begin{equation}
\frac{F_{\rm el}^{\rm DH}}{TV}=
-\frac{\kappa^3}{12\pi}\tau^{\rm\scriptscriptstyle DH}(\kappa a)
\label{FelDH}
\end{equation}
Here the inverse screening length $\kappa$ is given by
$\kappa^2=4\pi q^2l_B(\rho_++\rho_-)$, the Bjerrum lenght
$l_B=e^2/\varepsilon T$ and
\begin{equation}
\tau^{\rm\scriptscriptstyle DH}(x)=
\frac{3}{x^3}\left(\ln(1+x)-x+\frac{x^2}{2}\right)
\label{tauDH}
\end{equation}

Clearly, Eq. \eqref{Ur} for $U(r)$ can not be used in the RPA
approach if $\kappa a\sim1$. As the value of $\kappa a$ increases
the hard-sphere interactions between the charges  become more and
more important and they should be taken into account. 
For this purpose we introduce the following form of $U(r)$
\begin{equation}
U(r)=\frac{q^2}{r\varepsilon}(1-\exp(r/a))
\label{Ur_cut}
\end{equation}
which suppresses electrostatic interactions on the scale of the
size of the monomer. The Fourier transform of the above potential
can be easily calculated
\begin{equation}
U(k)=\frac{4\pi}{k^2(1+k^2a^2)}
\label{Uk}
\end{equation}
and for the electrostatic free energy given by Eq. \eqref{Fel} we get
\begin{eqnarray}
\frac{F_{\rm el}}{TV}=\frac{1}{4\pi^2}\int\limits_0^{\cal K}
\left[\ln\left(1+\frac{\kappa^2(1+g(k))}{2k^2(1+k^2a^2)}\right)-\right.\nonumber\\
\left.\frac{\kappa^2}{k^2(1+k^2a^2)}\right]k^2dk
\label{Felcut}
\end{eqnarray}
The integral in Eq. \eqref{Felcut} can be done analytically for
the case of $N=1$ and $\cal K=\infty$ which results in the
following expression for the free energy
\begin{equation}
\frac{F_{\rm el}^{\infty}}{TV}=-\frac{\kappa^3}{12\pi}\tau^{\infty}(\kappa a)
\label{FelHS}
\end{equation}
where
\begin{equation}
\tau^{\infty}(x)=\frac{1}{x^3}\left(-1+\frac{3x^2}{2}+\frac{1+x-2x^2}{(1+2x)^{1/2}}\right)
\label{tauHS}
\end{equation}

Another way to account for the hard sphere nature of the charges is to 
use the potential \eqref{Ur} but cut integration in \eqref{Fel} at ${\cal K}=1/a$.
This results in the following form of the free energy
\begin{equation}
\frac{F_{\rm el}^{\cal K}}{TV}=
-\frac{\kappa^3}{12\pi}\tau^{\scriptscriptstyle\cal K}(\kappa a)
\label{FelKA}
\end{equation}
where
\begin{equation}
\tau^{\cal\scriptscriptstyle K}(x)=\frac{1}{\pi}
\left(2\arctan(x^{-1})-\frac{\ln(1+x^2)-x^2}{x^3}\right)
\label{tauKA}
\end{equation}

\begin{figure}
\includegraphics[width=8cm]{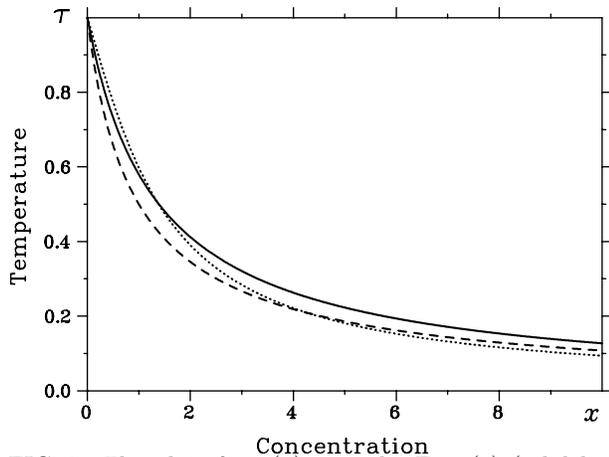}
\caption{The plots for $\tau(x)$  
given by Eq. \eqref{tauDH} (solid line), 
Eq. \eqref{tauHS} (dash line), and Eq. \eqref{tauKA} (dot line). }
\label{tau_plots}
\end{figure}

\begin{figure}
\includegraphics[width=8cm]{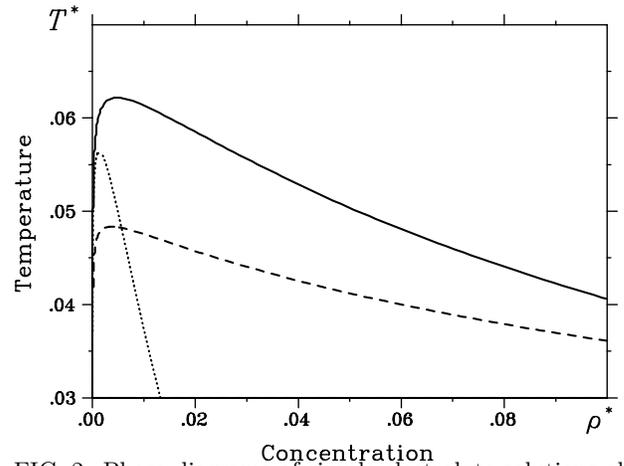}
\caption{Phase diagrams of simple electrolyte solutions obtained using different 
forms of electrostatic free energy given by Eq. \eqref{FelDH} (solid line),
Eq. \eqref{FelHS} (dash line), and Eq. \eqref{FelKA} (dot line). }
\label{diagrams-simple}
\end{figure}

In Figure \ref{tau_plots} we compare the results for 
$\tau^{\rm\scriptscriptstyle DH}(x)$, 
$\tau^{\infty}(x)$, 
and $\tau^{\cal\scriptscriptstyle K}(x)$, 
given by Eqs. \eqref{tauDH}, \eqref{tauHS}, and \eqref{tauKA} respectively. 
The curves presented in the figure show similar behavior as parameter
$x=\kappa a$ increases. In Figure \ref{diagrams-simple} we also show the phase 
diagrams  obtained on the plane $(\rho^*,T^*=a/l_B)$ using different form of the
electrostatic free energy given by Eqs. \eqref{FelDH}, \eqref{FelHS}, and 
\eqref{FelKA}. The figure clearly shows that the phase diagram obtained using Eq. 
\eqref{FelHS} is in reasonable agreement with Debye-H$\rm\ddot u$ckel theory 
where as the one obtained using Eq. \eqref{FelKA} differs significantly.
This fact leads us to the conclusion that
the hard-core interactions between the charges could be correctly taken 
into account through the potential $U(r)$ given by Eq. \eqref{Ur_cut}, but not 
through the large $k$ cut-off integration parameter ${\cal K}=1/a$. However, 
in the next section we show that a different form of the cut-off $\cal K$ is very 
important for the proper discription of {\it polyelectrolyte} solutions.

\section{Polyelectrolyte Solutions}
We first examine the dilute case of infinitely long
polyelectrolyte chains. In order to calculate the integral
\eqref{Fel} analytically we assume that the macromolecules obey
Gaussian statistics. For $g(k)$ defined by Eq. \eqref{gkdef} we
use the following approximation
\begin{equation}
g(k)=1+\frac{N}{1+Nb^2k^2/12}\simeq 1+\frac{12}{b^2k^2}
\label{gk_app}
\end{equation}
where $b^2$ is the mean-squared distance between the neighboring
monomers of the chain. For simplicity we assume here that $b=a$.
We also use $U(r)$ in the form \eqref{Ur} valid for the dilute
case $\kappa a\ll1$ and we rewrite Eq. \eqref{Felcut} with $\cal
K=\infty$ in the form
\begin{equation}
\frac{F_{\rm el}}{TV}=\frac{1}{4\pi^2}\int\limits_0^{\infty}
\left[\ln\left(1+\frac{\kappa^2}{k^2}
+\frac{6\kappa^2}{a^2k^4}\right)-\frac{\kappa^2}{k^2}\right]k^2dk
\label{FelNinfty}
\end{equation}
Evaluation of the above integral leads to
\begin{equation}
\frac{F_{\rm el}}{TV}=
\frac{\kappa^{3/2}}{12\pi a^{3/2}}\tau(\kappa a)
\label{FelNinfty1}
\end{equation}
where
\begin{equation}
\tau(x)=\frac{12-\sqrt{6}x-x^2}{(\sqrt{24}+x)^{1/2}}
\label{tauNinfty}
\end{equation}
The result $F_{\rm el}\sim\kappa^{3/2}$ for infinitely long
Gaussian chains was previously obtained in a number of papers (for
example see Refs. \CITE{Borue88,Olvera95}).

With the use of Eq. \eqref{FelNinfty1} we take the second derivative of the
free energy \eqref{F} with respect to the density
$\rho=\rho_++\rho_-$ and in the limit of $\rho\to 0$ we find
\begin{equation}
\frac{\partial^2}{\partial\rho^2}\left(\frac{F_{\rm el}}{TV}\right)\simeq
\frac{1}{2\rho}-\frac{\kappa^{3/2}}{64\pi b^{3/2}\rho^{2}}<0
\label{stabilityNinfty}
\end{equation}
The negative sign of the second derivative of the free energy
shows that dilute solution of infinitely long polyelectrolyte
chains are unstable in good solvent conditions.\cite{Mahdi00} In
Figure \ref{diagrams-gauss} we show the phase diagrams on the
plane $(\rho^*,T^*=a/l_B)$  obtained for different chain lengths
$N$. The electrostatic contribution to the free energy used to
obtain the diagrams is
\begin{eqnarray}
\frac{F_{\rm el}}{TV}=\frac{1}{4\pi^2a^3}\int\limits_0^{\infty}
&&\left[\ln\left(1+\frac{(\kappa a)^2(3/2+Nk^2/12)}{k^2(1+k^2)
(1+Nk^2/12)}\right)-\right.\nonumber\\
&&\left.\frac{(\kappa a)^2}{k^2(1+k^2)}\right]k^2dk
\label{FelN}
\end{eqnarray}
and it is evaluated numerically. The figure clearly demonstrates
that as $N$ increases the region of instability becomes broader
and broader even at very dilute concentrations. The same results
can be obtain if we introduce a constant cut-of value 
${\cal K}^*={\cal K}a=2\pi$.

We suggest here that the unphysical behavior described above
arises from the inappropriate contribution to the RPA
electrostatic free energy at large $k$. At large $k$ values the
chains are perturbed due to the electrostatic interactions. RPA,
however, is only valid if the interactions do not modify the chain
conformations. Therefore, the cut-off has to be chosen such that
we only include the electrostatic contributions that can be
accounted by RPA. It is straight forward to determine the cut-off
given that the major feature of polyelectrolyte solutions is the
existence of a concentration dependent correlation length
$\xi$.\cite{deGennes79} On the scales smaller then $\xi$ the chain
is stretched and on larger scales it obeys Gaussian statistics.
For strongly charged polyelectrolytes $\xi$ can be estimated
as\cite{Schiessel,Dobrynin99}
\begin{equation}
\xi=\frac{1}{(\rho^*)^{1/2}}
\label{ksi}
\end{equation}
In Fourier space scales smaller than $\xi$ are given by
$ka>2\pi/\xi$. Since the contributions to the free energy on these
scales can not be accounted within the RPA approach, we cut
integral \eqref{FelN}
\begin{equation}
{\cal K}^*=2\pi(\rho^*)^{1/2}
\label{calK2}
\end{equation}

\section{Results, Discussion and Conclusions}

\begin{figure}
\includegraphics[width=8cm]{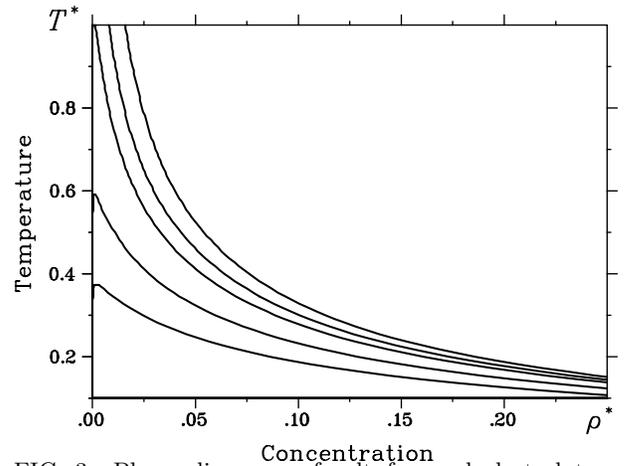}
\caption{Phase diagrams of salt free polyelectrolyte solutions
obtained  within classical RPA approach for different lengths of the polymers $N$.
From buttom to top $N=10,20,50,100,1000$.  }
\label{diagrams-gauss}
\end{figure}

\begin{figure}
\includegraphics[width=8cm]{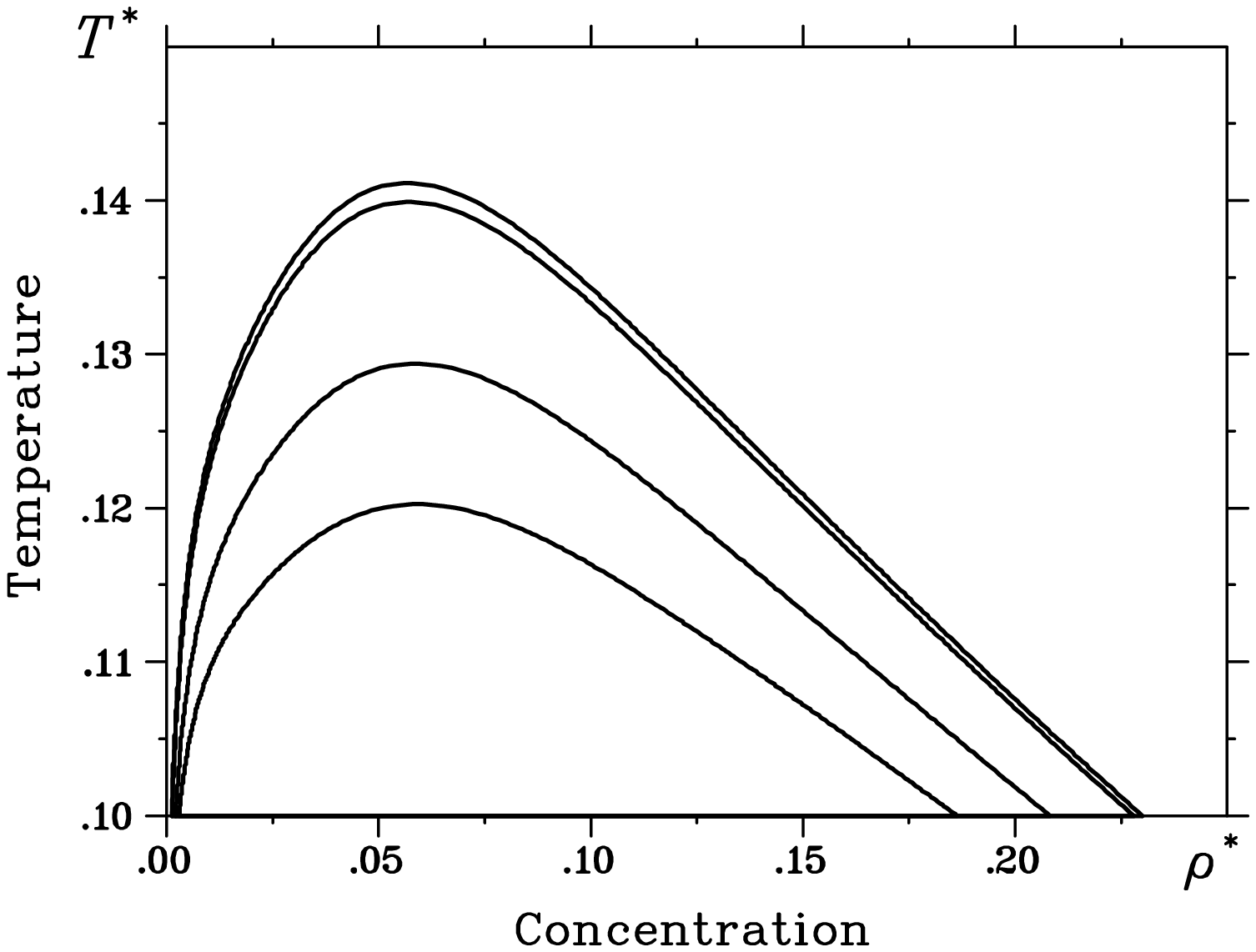}
\caption{Phase diagrams of salt free polyelectrolyte solutions
obtained  within modified RPA approach 
with ${\cal K}^*$ given by Eq. \eqref{calK2}
for different lengths of the polymers $N$.
From buttom to top $N=50,100,1000,10000$.  }
\label{diagrams}
\end{figure}

In Figure \ref{diagrams} we show the modified RPA phase diagrams
obtained by evaluating the integral \eqref{FelN} up to $ka={\cal K}^*$
with ${\cal K}^*$ given in \eqref{calK2}. The modified RPA phase
diagram has a much lower value of the critical temperature for
large $N$ than the standard RPA phase diagram Figure
\ref{diagrams-gauss}. The critical temperature initially increases
as $N$ increases in the modified RPA phase diagram, and it
plateaus at very high $N$ values at about $l_B/b \sim 7.3$. Since
most polymers have values of $l_B/b < 4.2$ at room temperature,
salt-free solutions of polyelectrolytes with monovalent
counterions are water soluble at room temperatures in agreement
with the experiments.

Another important feature of the modified RPA phase diagram in
Figure \ref{diagrams} is the prediction that the critical
concentration remains nearly constant as $N$ increases in
agreement with the simulations. This is only true if the
contributions from fluctuations of $k>\cal K$ are completely
ignored. That is, if we add the RPA electrostatic contribution
from the $2 \pi/a > k > \cal K$ in the free energy using the
structure function for rod like units, though the critical
temperature only increases slightly, the critical concentration
goes to zero as $N$ increases. This suggests that the charge
fluctuations at lower length scales do not contribute to the free
energy or that they are strongly suppressed. The suppression of
charge fluctuations on shorter length scales that $\xi$ can be
explained by the fact that the electrostatic interactions are not
screened at these length scales. Therefore, the counterions and
monomers are strongly correlated. Indeed, these correlations
stretched the chains.

We argue above that in polyelectrolyte solutions the cut-of is a
natural inverse length over which RPA breaks down. It is important
to point out, however, that a concentration dependent integration
cut-off is also obtained in the one component
plasma,\cite{Brilliantov02} in which counterions fluctuate over a
non-fluctuating charged medium. The physical reason and the
concentration dependence of the cut-off in the one component
plasma explained in Ref. \CITE{Brilliantov02} is different than in
polyelectrolytes.  In a one component plasma the integration in
the electrostatic free energy given by the RPA approach should be
carried  over a finite number of the wave-vectors $\bf k$ in the
same way as in the Debye theory of the specific heat of
solids.\cite{Ziman64} Namely, the total number of degrees of
freedom in the system $3{\cal N}=3\rho V$ should be equal to the
total number of physically different modes with wave-vectors $\bf
k$ within the spherical shell of radius $\cal K$. The number of
modes is twice the number of the wave-vectors since each $\bf k$
has a sine a cosine mode. Therefore, one obtains
\begin{equation}
2V\int\limits_0^{\cal K}\frac{d^3k}{(2\pi)^3}=3{\cal N}
\label{modes}
\end{equation}
which leads to
\begin{equation}
{\cal K}^*=(9\pi^2\rho)^{1/3}
\label{calK1}
\end{equation}

The above argument assumes that the important length scale is the
average distance between ions $d \sim 2 \pi/{\cal K}\sim
\rho^{-1/3}$, and that this length scales imposes a "periodicity"
even though the non fluctuating support for the ions is a solid
with no structure.

The distance between charges $d \sim 1/\rho^{1/3}$ is an important
length scale to determine the break down of DH in simple
electrolyte solutions.\cite{Brilliantov98} Therefore, in simple
electrolytes $d$ is an important length scale as the system gets
concentrated when the electrostatic energy between two charges
separated by $d$, $e^2/(4 \pi \varepsilon d) > K_B T$ or $l_B/d >
1$. Interestingly, the integration to a concentration dependent
cut-off ${\cal K}\sim \rho^{1/3}$ versus integrating to the inverse
hard core size of the ions ${\cal K}\sim 1/a$ does not affect
significantly the RPA phase diagrams in simple electrolytes.
Instead, in salt-free polyelectrolyte solutions a concentration
dependent cut-off is essential to recover reasonable values of the
critical temperature as $N$ increases and a $N$ independent
critical concentration.

\begin{figure}
\includegraphics[width=8cm]{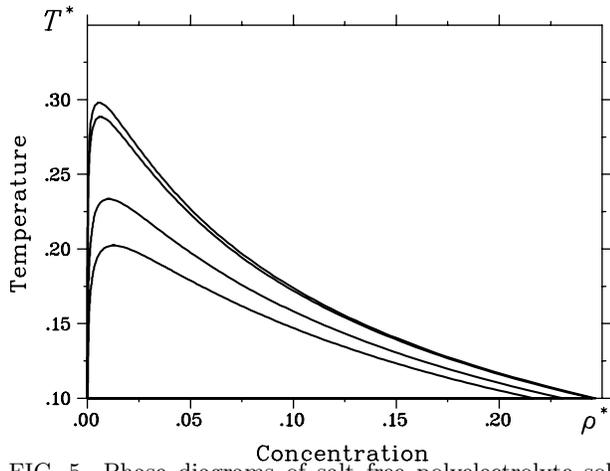}
\caption{Phase diagrams of salt free polyelectrolyte solutions
obtained  within modified RPA approach with ${\cal K}^*$ given by Eq. 
\eqref{calK1} for different  lengths of the polymers $N$.
From buttom to top $N=50,100,1000,10000$.  }
\label{diagrams-d}
\end{figure}

Though we argue that in salt-free polyelectrolyte solutions the
natural length scale that determines the cut-off is $\xi$, we
obtain very similar phase diagrams if we instead use $d$ as shown 
in Figure \ref{diagrams-d}. Notice
that the importance of a concentration dependent cut-off is at
dilute solutions where $\xi
> d$. Therefore, there should be counterion fluctuations at length scales
$d<l<\xi$ that we did not include in our RPA approach. If we add a
RPA contribution from these counterion fluctuations (only the
counterions) we get negligible corrections to the phase diagram.
Our arguments are in agreement with the results in Ref.
\CITE{Donley97} where a self-consistent polyelectrolyte RPA where
the screening length from polyions is wavevector-dependent due to
chain connectivity shows that at high wavevectors (short length
scales), only counterions contribute to screening. At lower
wavevectors, both polyions and counterions contribute to
screening.

In summary, we show that in salt-free solutions a wave vector
dependent cut-off is essential to recover physical phase diagrams.
The cut-off is determined by the applicability limit of the RPA.
In the presence of salt the electrostatic contribution to the free
energy reduces to the well known DH limiting law.\cite{Borue88} In
that case reasonable phase diagram are obtain without a
concentration dependent cut-off,\cite{Mahdi} although in principle
one should only include fluctuations from the monomers on the
length scale where the RPA approach is valid (i.e., larger than
$\xi$).

We acknowledge the financial support of the NIH grant number
GM62109-02 and of the Institute for Bioengineering and
Nanosciences in Advanced Medicine (IBNAM) at Northwestern
University.

\end{document}